# Automated Work Records for Precision Agriculture Management: A Low-Cost GNSS IoT Solution for Paddy Fields in Central Japan.

Malte Grosse[1, *], Kiyoshi Honda[1,2], Cornelius Specht [3], Juan Cesar Pineda [2]

1) Chubu University, Japan
 ts22851-7392@sti.chubu.ac.jp
2) ListenField Inc., Japan
3) Media University Stuttgart, Germany

* Corresponding author

**ABSTRACT**

Agricultural field operations are generally tracked as work records (WR), incorporating data points such as; work type, machine type, timestamped trajectories and field information. WR data which is automatically recorded by modern machinery equipped with Information and Communication Technologies (ICT) can enable efficient farm management decision making. Globally, farmers often rely on aged or legacy farming machinery and manual data recording, which introduces significant labor costs and increases the risk of inaccurate data input. To address this challenge, a field study in Central Japan was conducted to showcase automated data collection by retrofitting legacy farming machinery with low-cost Internet of Things (IoT) devices. For single-purpose vehicles (SPV), which only carry out single work types such as planting, LTE (Long Term Evolution) and Global Navigation Satellite System (GNSS) units were installed to record trajectory data. For multi-purpose vehicles (MPV), such as tractors which perform multiple work types, the configuration settings of these vehicles had to include implements and attachments data. To obtain this data, industry standard LTE-GNSS Bluetooth gateways were fitted onto MPV and low-cost BLE (Bluetooth Low Energy) beacons were attached to implements. After installation, over a seven month field preparation and planting period 1,623 WR, including 421 WR for SPV and 1,120 WR for MVP, were automatically obtained. For MPV, the WR included detailed configuration settings enabling detection of the specific work types. These findings demonstrate the potential of low cost IoT GNSS devices for precision agriculture strategies to support management decisions in farming operations.

## INTRODUCTION
Smart Farming has gained prominence due to the increase of efficiency of farmers in precise agricultural management through the utilization of Information and Communication Technology (ICT) (Nanseki, Li, & Chomei, 2023). Agricultural field operations data, generally called work records (WR), incorporate data points such as; work type, machine type, timestamped trajectories and field information, which support precision agriculture application and improve farm management decision making (Topakci et al., 2010).
However, many farmers worldwide still rely on aged machinery and manual data recording methods, leading to high labor costs and potential errors (Jiang et al., 2017). To address this challenge, a field study conducted in Central Japan, demonstrating the integration of low-cost Internet of Things (IoT) technologies with legacy farming machinery to automate WR collection.

## MATERIALS AND METHODS
This study was carried out on a farm located in the Chita area, Aichi Prefecture Japan. Three farmers utilize together 10 agricultural machines to cultivate over 500 paddy fields covering 64 hectares. Two types of farming vehicles were retrofitted with IoT devices: single-purpose vehicles (SPV) and multi-purpose vehicles (MPV). For SPV, such as harvesters, planters and sprayers, Long Term Evolution (LTE) mobile communication and Global Navigation Satellite System (GNSS) units were installed to record timestamped trajectory data. For MPV like tractors, industry-standard LTE-GNSS Bluetooth gateways were fitted onto the vehicles, while low-cost BLE beacons were attached to implements to capture detailed configuration settings enabling specific work type detection. The BLE beacons with five-year battery life spans were attached to 17 implements such as; grass cutters, harrowers, seeders, plow and rotary tillers.

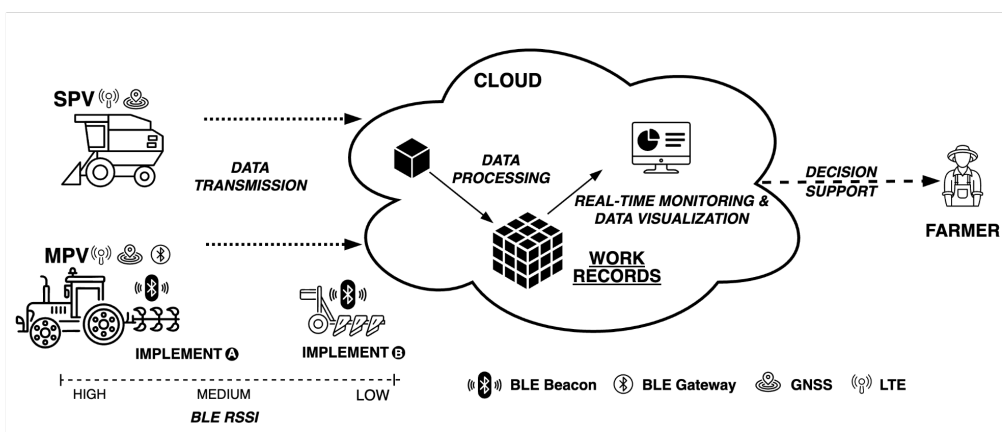

**Figure 1. Automated Data-Collection for Work Records using BLE**

As seen in Figure 1, the installed IoT devices automatically record and transmit farming machinery work data to a cloud-based platform for storage, processing, and real-time monitoring.
SPVs send location data directly to the cloud. MPV, which are installed with BLE gateways, actively scan for surrounding BLE beacons, allowing location data to be associated with beacon information such as Unique Identifiers (UID) associated with specific implements and Received Signal Strength Indicators (RSSI) to be transmitted to the cloud.
Further data processing is done in the cloud such as adding field information, machinery data and the identification of configuration settings using the UID. To mitigate WR inaccuracies, a set of algorithms is applied during data processing, which identifies and adjusts for parked machinery on fields, disrupted work operations, navigational transitions between adjacent fields and proximity of non-attached implements.

## RESULTS AND DISCUSSION
During the seven-month field study period, 1,623 WR, including 421 WR for SPV and 1,120 WR for MVP, were automatically obtained. For MPV, the captured WR included detailed configuration settings, allowing for specific work type detection.
Comparison of automatically collected WR data against manually recorded WR data shows discrepancies in work dates and existence of records (errors which farmers indicated is due to the labor-intensive nature of manual record keeping).

Additionally, subsequent analysis of automatically generated records revealed operational challenges due to inefficient routing of machinery between fields.

As a side effect, the obtained trajectory data was utilized to digitize field boundaries thus assisting farmers with better farm management system information.

To make the WR data accessible to farmers, a web-based application was developed, which visualizes machine type, work trajectory, work duration, work type and field information (Figure 2.). The result of this enabled real-time data monitoring and collaborative work within the team of farmers as well as optimizing their farming operations.

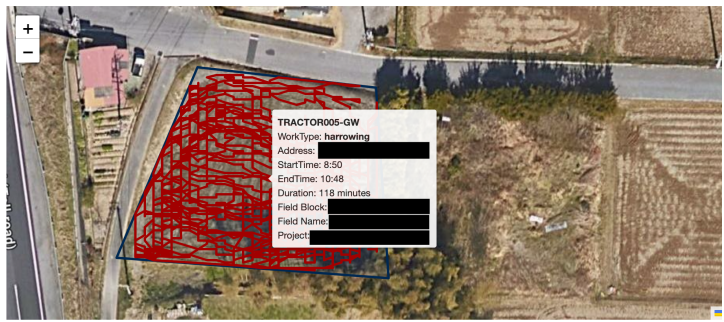

**Figure 2. Application visualizing work record data**

## CONCLUSIONS

The deployment of low-cost IoT devices enabled the capture of comprehensive WR data without relying on manual input.

This study successfully demonstrated that retrofitting legacy farming machinery with low-cost IoT devices can automate work record collection, showcasing the potential of this approach to replace manual data recording methods. By enabling farmers to monitor specific work patterns, and make decisions based on real-time data, the system has the capability to improve operational efficiency and contribute to agricultural productivity gains.

Future work should extend beyond seven months to cover the entire crop cycle from planting to harvesting, enabling for example the calibration of crop models as these rely on detailed and correct WR data.

Finally, integrating this data with farm management information systems (FMIS) via standards such as the Japanese NoukiOpenAP/農機OpenAPI (NARO, 2024) could provide interoperability between different platforms, further facilitating informed decision-making in agriculture.

## ACKNOWLEDGEMENT

This work was supported by JST SPRING, Grant Number JPMJSP2158.

## REFERENCES.